\documentclass[prd,a4paper,twocolumn,preprintnumbers,nofootinbib,superscriptaddress]{revtex4}

\pdfoutput=1
\usepackage[english]{babel}
\usepackage{amsmath,amssymb,amsfonts, bm,bbm,slashed, subdepth}
\usepackage{graphicx}
\usepackage{hyperref}
\usepackage{enumerate}
\usepackage{setspace}
\usepackage{booktabs, tabularx}
\usepackage{units}
\usepackage{color}
\usepackage{multirow}
\usepackage[dvipsnames]{xcolor}
\usepackage[pscoord]{eso-pic}
\usepackage[normalem]{ulem}

\newcommand{\be}{\begin{equation}}
\newcommand{\ee}{\end{equation}}
\newcommand{\ba}{\begin{array}}
\newcommand{\ea}{\end{array}}
\newcommand{\bea}{\begin{eqnarray}}
\newcommand{\eea}{\end{eqnarray}}
\newcommand{\balg}{\begin{align}}
\newcommand{\ealg}{\end{align}}
\newcommand{\bit}{\begin{itemize}}
\newcommand{\eit}{\end{itemize}}

\newcommand*\diff{\mathop{}\!\mathrm{d}}

\makeindex

\begin{document}
\preprint{ULB-TH/20-02}

\title{Dark matter as a heavy thermal hot relic}

\author{Thomas Hambye}
\email{thambye@ulb.ac.be}
\affiliation{Service de Physique Th\'eorique, Universit\'e Libre de Bruxelles, Boulevard du Triomphe, CP225, 1050 Brussels, Belgium}

\author{Matteo Lucca}
\email{mlucca@ulb.ac.be}
\affiliation{Service de Physique Th\'eorique, Universit\'e Libre de Bruxelles, Boulevard du Triomphe, CP225, 1050 Brussels, Belgium}

\author{Laurent Vanderheyden}
\email{lavdheyd@ulb.ac.be}
\affiliation{Service de Physique Th\'eorique, Universit\'e Libre de Bruxelles, Boulevard du Triomphe, CP225, 1050 Brussels, Belgium}

\begin{abstract}
If, during the early Universe epoch, the dark matter particle thermalizes in a hidden sector which does not thermalize with the Standard Model thermal bath, its relativistic thermal decoupling can easily lead to the observed relic density, even if the dark matter particle mass is many orders of magnitude heavier than the usual $\sim$ eV hot relic mass scale. This straightforward scenario simply requires that the temperature of the hidden sector thermal bath is one to five orders of magnitude cooler than the temperature of the Standard Model thermal bath. In this way the resulting relic density turns out to be determined only by the dark matter mass scale and the ratio of the temperatures of both sectors. In a model independent way we determine that this can work for a dark matter mass all the way from $\sim 1$ keV to $\sim 30$ PeV. We also show how this scenario works explicitly in the framework of two illustrative models. One of them can lead to a PeV neutrino flux from dark matter decay of the order of the one needed to account for the high energy neutrinos observed by IceCube.
\end{abstract}

\maketitle

To account for the fact that dark matter (DM) is res\-ponsible for about $\sim26\,\%$ of the energy content of the Universe today, the thermal non-relativistic freeze-out scenario is particularly straightforward. It is based on the simple assumption that the DM particles interact in a sizeable way with some of the Standard Model (SM) particles, so that it thermalizes with them during the radiation dominated early Universe epoch. The decoupling from thermal equilibrium it involves can easily lead to the observed relic density for a DM mass scale all the way from few keV to $\sim 100$~TeV \cite{Griest:1989wd}. As well known, this scenario is quite different from the case of a relativistic DM decoupling, which leads to a suppressed enough relic density  only if the DM mass is around a scale as low as the $\sim 10$~eV scale. This possibility is however excluded by large scale structure (LSS) formation constraints, see e.g.~\cite{Viel:2013apy, Safarzadeh:2018hhg, Banik:2019smi}.

This Weakly Interacting Massive Particle (WIMP) miracle remains as of today the most straightforward, and in many ways the most attractive, possibility to account for the DM relic density. However, the non-observation of any of the various WIMP candidates which should have been already observed, in particular by direct detection experiments, challenges this explanation, even if many other WIMP candidates are still perfectly viable. Furthermore, the non-observation of any new physics around the TeV scale at the LHC raises some doubts on the fact that there could be any new fundamental physics associated to this scale at all. The new physics scale could be many orders of magnitude above the LHC scale, or much below if this new physics only couples feebly to the SM particles.

All this motivates searches for new DM production mechanisms at scales much higher than the LHC scale \cite{Chung:1998zb,Kolb:1998ki,Randall:2015xza,Berlin:2016vnh,Berlin:2016gtr,Harigaya:2016nlg,Tenkanen:2016jic,Berlin:2017ife,Bramante:2017obj,Kolb:2017jvz,Blanco:2017sbc,Cirelli:2018iax,Davoudiasl:2019xeb,Kim:2019udq,Heurtier:2019eou,Baker:2019ndr,Chway:2019kft,Heurtier:2019beu} and/or with the new physics coupling only feebly to the SM. The possibility that the DM particle mass could be much above the TeV scale is also interesting for various other reasons. An experimentally motivated example is that PeV DM decay could constitute a viable explanation for the $\sim$~PeV neutrinos events observed by the IceCube experiment \cite{Haack:2017dxi, Kopper:2017zzm, Aartsen:2018mxl}.

In the following we consider a scenario where the DM does not thermalize with the SM thermal bath, but still thermalizes with other particles within a hidden sector. Thus, the DM couples only feebly to the SM particles but couples  much more to the other particles in the hidden sector. In this case, the temperature of the hidden sector $T'$ is expected to be different from the temperature of the SM bath, $T$. Here, for the purpose of DM production, the $T'/T$ ratio is an initial condition, presumably fixed by inflation physics. For such a setup, the possibility that DM undergoes a non-relativistic freeze-out within the hidden sector (i.e.~that it decouples while being already largely non-relativistic) has been considered in a long series of works, see e.g.~\cite{Feng:2008mu,Chu:2011be, Berlin:2016vnh, Berlin:2016gtr,Arcadi:2019oxh}. However, the possibility that it decouples when it is still relativistic, or close to relativistic, has, to our knowledge, been hardly discussed, even though it is a particularly straightforward and somewhat self-evident scenario. It has been considered as a possibility to have a $\sim$~keV warm DM candidate with $T'/T$ slightly below unity \cite{Colombi:1995ze,Bode:2000gq,Viel:2005qj} (e.g.~obtained through decoupling of relativistic species, starting from $T'/T=1$ at a high temperature). In \cite{Hambye:2019tjt} it has been briefly discussed to show that this scenario is viable  for $T'/T\ll 1$ (and large DM masses).

Here we consider this regime in details. In particular we determine the range of DM masses it allows, and we also consider 2 explicit models to illustrate how this scenario works in concrete setups. The first model assumes a heavy mediator, whereas the second one does not involve any mediator. For both cases, we show that relativistic or almost relativistic thermal decoupling within the hidden sector can straightforwardly lead to the observed relic density for DM masses much larger than in the ordinary freeze-out scenario with a single thermal bath. The relativistic decoupling regime is in particular the regime which turns out to allow the highest DM masses. It allows masses up to several tens of PeV. Such a scenario fits very well with the idea that there could be a whole new sector associated to DM, and nothing tells that this whole new sector should necessarily interact sizeably with the SM sector in other ways than gravitationally.

If DM decouples relativistically, its number density at time of decoupling is simply given by 
\begin{equation}
n_{DM}=\frac{\zeta(3)}{\pi^2}g_{DM}^{(n)} T^3\cdot \left(\frac{T'}{T}\right)^3\,,
\label{nDMrelativistic}
\end{equation}
where $g_{DM}^{(n)}$ is the effective number of degrees of freedom of the DM particle contributing to the DM number density (i.e.~$g_{DM}\cdot c$ with $g_{DM}$ as the number of DM degrees of freedom and $c=1,3/4$ for bosons and fermions respectively). The $(T'/T)^3$ factor arises if DM thermalizes within a thermal bath which has temperature $T'$, rather than the temperature $T$ of the SM thermal bath.

Well known examples of hot relics are the SM neutrinos. Since $T_\nu=T_\gamma$ when neutrinos decouple, they lead to a too large relic density unless they are very light, $\Omega_\nu\simeq 0.26\cdot (\sum m_\nu/10\,\hbox{eV})$. Neutrinos decouple relativistically because their annihilation processes into charged leptons are weak processes, i.e.,~mediated by the much heavier $W$ and $Z$ bosons. For $m_W \gg T \gg m_{\nu,l}$, the annihilation cross section is $\langle \sigma v \rangle \propto \alpha_W^2 T^2/m_{W}^4$. This gives a decoupling temperature $T^\nu_{dec}\simeq (m_W^4/m_{Planck} \alpha_W^2)^{1/3}\sim 1 \,\hbox{MeV}\gg m_\nu$ as solution of the decoupling condition $\Gamma_\nu=n_\nu \langle \sigma v \rangle=H$. 

Now for a DM particle which decouples within a thermal bath with  a temperature $T'$ smaller than $T$, one immediately understands that the $10$~eV scale is traded for a scale a factor $(T/T')^3$ times larger. Thus, the DM mass scale can be much larger, even though it is a hot relic. Explicitly, we can make use of Eq. \eqref{nDMrelativistic} to calculate
\begin{equation}
\Omega_{DM}= 1.74 \times 10^{11}\cdot \left(\frac{T'_{dec}}{T_{dec}}\right)^3\left(\frac{m_{DM}}{1\,\hbox{TeV}}\right) \left(\frac{g_{DM}^{(n)}}{g^S_\star(T_{dec})}\right),
\label{OmegaDMrelativistic}
\end{equation}
where $T_{dec}$ and $T'_{dec}$ are the values of $T$ and $T'$ when DM decouples, with the last factor accounting for the fact that the number of entropy degrees of freedom at decoupling is not the same as today. The other way around, this equation gives the $T'/T$ ratio one needs to account for the 26\% if DM has a given mass,
\begin{equation}
\frac{T'_{dec}}{T_{dec}}=1.14\times10^{-4}\cdot \left(\frac{1\,\hbox{TeV}}{m_{DM}}\right)^{1/3}  \left(\frac{g^S_\star(T_{dec})}{g_{DM}^{(n)}}\right)^{1/3}\,.
\label{TprimeoverT}
\end{equation}
As briefly discussed recently in \cite{Hambye:2019tjt}, this value of $T'/T$ constitutes a ``$T'/T$ floor'' in the sense that, thermally,  one cannot have more particles left than for the case of DM decoupling relativistically. Thus, for any smaller value of $T'/T$, the number of DM particles left is too small to account for the 26\%. Note that, as Eq.~(\ref{OmegaDMrelativistic}) shows, the relic density is determined by only 2 input parameters, $m_{DM}$ and $T'/T$. Unlike for the usual freeze-out scenario there is no dependence on the value of the annihilation cross section (and thus on the values of the masses and couplings determining this cross section).

Combining this $T'/T$ floor value with the lower bound on the mass of a sterile neutrino, coming from the requirement that it is enough non-relativistic at the time of LSS formation \cite{Viel:2013apy}
\begin{equation}
m_{DM}\gtrsim  3.3 \, \hbox{keV} \cdot \left(\frac{T'_{dec}}{T_{dec}}\right)\,,
\label{massLSSbound}
\end{equation}
one obtains the new lower bound on $m_{DM}$ 
\begin{equation}
m_{DM}\gtrsim  0.48 \, \hbox{keV} \cdot \left(\frac{g_\star^S(T_{dec})}{g_{DM}^{(n)}}\right)^{1/4}\,.
\label{massLSSboundbis}
\end{equation}
The extra $T'/T$ factor in Eq.~(\ref{massLSSbound}) accounts for the fact that the typical DM streaming speed at matter-radiation equality, which is the relevant quantity for LSS, scales as $v_s/c\propto T_{DM}/m_{DM}$, see e.g.~\cite{Bode:2000gq}.\footnote{Generalizing Eq.~(5) of \cite{Viel:2005qj} (see also \cite{Colombi:1995ze}) to arbitrary numbers of degrees of freedom gives the same result.} For this value of $m_{DM}$, Eq.~(\ref{TprimeoverT}) gives a value of $T'/T$ which  constitutes the maximal value of $T'/T$ along which DM can be a hot relic in agreement with LSS constraints. This gives
\begin{equation}
\frac{T'_{dec}}{T_{dec}}\lesssim 0.15\cdot \left(\frac{g_\star^S(T_{dec})}{g_{DM}^{(n)}}\right)^{1/4}\,.
\end{equation}

As for the upper bound on $m_{DM}$, since the decoupling is relativistic rather than non-relativistic, one does not need an annihilation cross section as large as for the non-relativistic case, i.e.~able to keep DM thermalized even when its number is already largely Boltzmann suppressed. Thus one can anticipate that the unitarity bound on the cross section allows much larger DM masses. But still a bound exists because one has to make sure that, in this scenario, DM has thermalized within the hidden sector and the larger $m_{DM}$, the smaller $T'/T$ must be, the smaller is the number of DM particles, the less they thermalize.
The condition to satisfy here is approximately that $\Gamma/H=n_{DM}^{eq}(T') \langle \sigma_{ann.}v\rangle/H>1$ holds at some point, taking into account that the annihilation cross section is bounded from above by unitarity. More explicitly, the condition $\Gamma/H=n_{DM}^{eq}(T') \langle \sigma_{ann.}v\rangle/H>1$ gives
\begin{equation}
\langle \sigma v_{rel}\rangle >\frac{1.67\pi^2 \sqrt{g_{\star}^{eff}(T_{dec})}}{\zeta(3) \cdot g_{DM}^{(n)}}\cdot \frac{T^2_{dec}}{T'^3_{dec}\,M_{Pl}}.\label{thermalizationconditiongeneral}
\end{equation}
This bound is to be combined with the unitarity constraint which holds on the cross section \cite{Griest:1989wd}
\begin{equation}
\sigma_{ann.}v_{rel}<\frac{\pi (2J+1)}{p^2_{DM}}v_{rel}=\frac{\pi(2J+1)}{m^2_{DM}}\frac{(1-v_{rel}^2/4)}{v_{rel}/4},
\label{unitaritygeneral}
\end{equation}
where $v_{rel}$ is the relative velocity between both annihilating DM particles and $J$ is the angular momentum between the in-going particles. This gives the following unitarity upper bound on $m_{DM}$,
\begin{equation}
m_{DM}\lesssim 30.3 \hbox{ PeV}\,\cdot(2J+1)^{3/5}.
\label{approxunitbound}
\end{equation}
Concretely, to get this bound one has to compute the maximum possible thermally averaged cross section $\left\langle\sigma v\right\rangle$ by integrating Eq.~(\ref{unitaritygeneral}) 
over all possible velocities\begin{eqnarray}
\left\langle \sigma v\right\rangle \equiv \frac{\int\sigma_{ann.} v_{rel} f_{v}(E_{1})f_{v}(E_{2})\diff p_{1}^{3}\diff p_{2}^{3}}{\int f_{v}(E_{1})f_{v}(E_{2})\diff p_{1}^{3}\diff p_{2}^{3}},\label{thermalaverage}
\end{eqnarray}
where 1 and 2 refer to the two in-going particles of energy $E_{1,2}$ and momentum $\vec{p}_{1,2}$. The velocity distribution, $f_{v}$, can be either the Fermi-Dirac or the Bose-Einstein distribution depending on the spin of the incident particles. This gives
\begin{equation}
\left\langle\sigma v\right\rangle<\frac{\pi (2J+1)}{4m_{DM}^2}x'^2\mathcal{I}(x';\epsilon),
\label{unitaritygeneralaverage}
\end{equation}
where $x'\equiv m_{DM}/T'_{dec}$, with $\epsilon = \pm 1$ for a fermion or a boson respectively, and where $\mathcal{I}$ is a (numerically computed) factor of order unity (for relativistic decoupling, $x'<1$) given by\footnote{One can check that $\mathcal{I}(x'\leqslant 1;1)\simeq 1.66$, $\mathcal{I}(1;-1)\simeq 1.7$ and $\mathcal{I}(x'<1;-1)\simeq 3.7$.}
\begin{eqnarray}
\mathcal{I}(x';\epsilon)\equiv &&\hspace{-0.5cm}\frac{1}{N^{2}}\cdot \int_{4x'^{2}}^{\infty}\diff w\int_{\sqrt{w}}^{\infty}\diff k_{+}\int_{-k_{-,max}}^{k_{-,max}}\diff k_{-}\nonumber\\
&&\hspace{-0.5cm}\times\left\lbrace\frac{\sqrt{w/(w-4x'^{2})}}{\left(e^{\frac{k_{+}+k_{-}}{2}}+\epsilon\right)\left(e^{\frac{k_{+}-k_{-}}{2}}+\epsilon\right)}\right\rbrace.
\end{eqnarray}
with
$N\equiv \int_{x'}^{\infty}\frac{\sqrt{k^{2}-x'^{2}}}{e^{k}+\epsilon}k \diff k$, $k_{\pm}\equiv (E_{1}\pm E_{2})/T_{dec}'$, $w\equiv s/T_{dec}'^{2}$,  $k_{-,max}\equiv \sqrt{1-4x'^{2}/w}\sqrt{k_{+}^{2}-w}$ and $s\equiv (p_1+p_2)^2$. 
Plugging Eq.~(\ref{TprimeoverT}) into Eqs. (\ref{thermalizationconditiongeneral}) and (\ref{unitaritygeneralaverage}), gives 
\begin{eqnarray}
m_{DM}< && \hspace{-0.5cm} 30.3 \hbox{ PeV}\,\cdot(2J+1)^{3/5}x'^{3/5}\left(\frac{\mathcal{I}(x';\epsilon)}{1.5}\right)^{3/5}\nonumber\\
&&\hspace{-0.5cm}\times\left(\frac{g_{\star}^{S}(T_{dec})}{100}\right)^{2/5}\hspace{-1.5mm}\left(\frac{100}{g_{\star}^{eff}(T_{dec})}\right)^{3/10}\hspace{-1.5mm}\left(g_{DM}^{(n)}\right)^{1/5} \,\hspace{-4mm}.
\label{unitarboundmodindep}
\end{eqnarray}
This bound is maximum when the relativistic decoupling occurs at the lowest $T'_{dec}$ value allowed in this regime, i.e.~$T'_{dec}\simeq m_{DM}$, leading to the roughly approximated unitarity bound of Eq.~(\ref{approxunitbound}). One concludes from this result that the relativistic decoupling scenario is perfectly viable for DM candidates around the PeV scale. 

Let us now analyse what this relativistic decoupling mechanism implies for the parameter space of two examples of explicit models.
The first model we consider is a minimal heavy mediator model where DM is a Dirac fermion annihilating into a pair of lighter Dirac fermions through a heavier 
real scalar $S$, $\psi_{DM}\overline{\psi}_{DM}\rightarrow S\rightarrow \psi'\overline{\psi}'$, as induced by the Lagrangian,
\begin{equation}
{\cal L} \owns - y_{DM} \overline{\psi}_{DM} \psi_{DM} S- y'\overline{\psi'} \psi' S\,.
\end{equation}
In this case, the annihilation cross section is given by $\langle \sigma_{ann.} v\rangle\simeq y_{DM}^2 y'^2 T'^2/(2\pi m_S^4)$ (for  the range $m_{DM}\lesssim T'\lesssim m_S$ which applies for a relativistic decoupling).
Solving the condition $\Gamma/H=n_{DM}^{eq}(T') \langle \sigma_{ann.}v\rangle/H=1$, one obtains
\begin{eqnarray}
T'_{dec}= && \hspace{-0.5cm} 19.5 \hbox{ TeV} \cdot \left(\frac{1}{\alpha\alpha'}\right)^{1/3}\left(\frac{m_{S}}{\text{PeV}}\right)^{4/3}\left(\frac{m_{DM}}{\text{10 TeV}}\right)^{2/9}\nonumber\\
&& \hspace{-0.5cm} \times\left(\frac{100}{g_{\star}^{S}(T_{dec})}\right)^{1/18}\left(\frac{1}{g_{DM}^{(n)}}\right)^{1/9},\label{eq:Tprimedec}
\end{eqnarray}
with $\alpha\equiv y_{DM}^2/4\pi$ and $\alpha'\equiv y'^2/4\pi$. For this model, the unitarity bound of Eq.~(\ref{unitarboundmodindep}) gives
\begin{equation}
m_{DM}\lesssim 38.0 \hbox{ PeV}\cdot x'^{3/5}\left(\frac{\mathcal{I}(x';1)}{1.5}\right)^{3/5},
\label{mDMexplicitA}
\end{equation}
using $J=0$, $g_{DM}^{(n)}=3$ and $g_\star^{eff}(T_{dec})=g_\star^s(T'_{dec})\simeq 106$. Using Eq.~(\ref{eq:Tprimedec}) in Eq.~(\ref{mDMexplicitA}), one also gets the following upper bound,
\begin{eqnarray}
\frac{m_{S}}{m_{DM}}< 922 \cdot \left(\alpha\alpha'\right)^{1/4}\left(\frac{\text{PeV}}{m_{DM}}\right)^{5/3}\left(\frac{\mathcal{I}(x';1)}{1.5}\right)^{3/4}.\,
\label{mSmDMupperbound}
\end{eqnarray}
Plugging the relativistic decoupling condition, ${T'_{dec}\gtrsim m_{DM}}$, in Eq. (\ref{eq:Tprimedec}), we also obtain
\begin{eqnarray}
&&\frac{m_{S}}{m_{DM}}\gtrsim 9.78 \cdot \left(\alpha\alpha'\right)^{1/4}\left(\frac{\text{PeV}}{m_{DM}}\right)^{5/12}.
\label{mSmDMlowerbound}
\end{eqnarray}

Note also that, considering the explicit form of the cross section, the unitarity bound requires that 
\begin{equation}
\frac{m_S}{T'_{dec}}> 2.15\cdot(\alpha\alpha')^{1/4}\left(\frac{1.5}{\mathcal{I}(x';1)}\right)^{1/4}.
\end{equation}
For perturbative couplings this is typically satisfied, as expected.

\begin{figure}[t]
	\centering
	\includegraphics[scale=1.2]{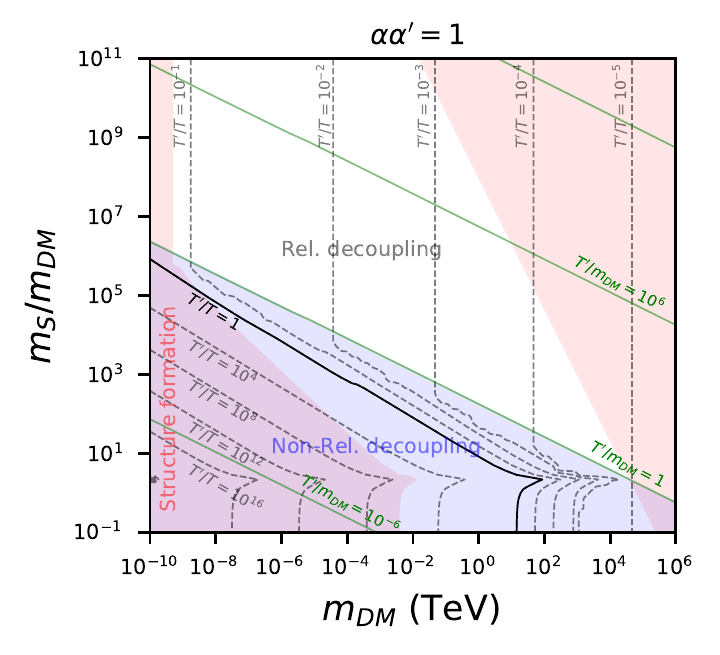}
	\includegraphics[scale=1.2]{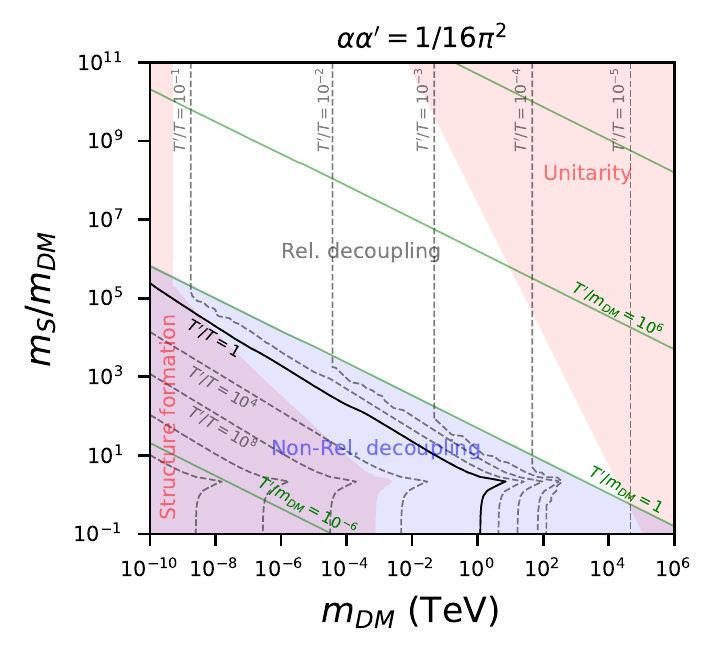}
	\caption{Allowed parameter space for two choices of the couplings, $\alpha\alpha '=1$ (top) and $\alpha\alpha '=1/16\pi^{2}$ (bottom). The regions excluded by the structure formation and unitarity constraints are shown in red, while the blue region indicates the non-relativistic regime. We also show various contours of $T'_{dec}/m_{DM}$ and $T'_{dec}/T_{dec}$.}
	\label{fig:Mass_to_mass}
\end{figure}

Fig.~\ref{fig:Mass_to_mass} summarizes the results obtained above by showing, for 2 values of $\alpha\alpha'$ and as a function of $m_{DM}$, the values of $m_S/m_{DM}$ excluded by the lower and upper bounds of Eqs.~(\ref{mSmDMlowerbound}) and (\ref{mSmDMupperbound}) respectively. In these figures we also show the regions excluded by LSS, Eq.~(\ref{massLSSboundbis}), as well as various isocontours of $T'_{dec}/T_{dec}$, Eq.~(\ref{TprimeoverT}), and $T'_{dec}/m_{DM}$. In the relativistic decoupling regime, the $T'_{dec}/T_{dec}$ isocontour lines are vertical because this ratio does not depend on $m_S$, Eq.~(\ref{TprimeoverT}). In Fig.~\ref{fig:Mass_to_mass} the unitarity absolute upper bound of Eq.~(\ref{unitarboundmodindep}), that is to say Eq.~(\ref{mDMexplicitA}), is saturated for $T'_{dec}/m_{DM}\simeq 1$ and $m_S/m_{DM}\sim 2$. For lower values of $m_{DM}$, a relativistic decoupling requires a mediator much heavier than the DM particle, $m_S\gg m_{DM}$. 

For lower values of $m_S/m_{DM}$, one enters in the non-relativistic decoupling regime (blue region in Fig.~\ref{fig:Mass_to_mass}), since this is the region for which $T'_{dec}/m_{DM}<1$. Here, the results of Fig.~\ref{fig:Mass_to_mass} follow from a slightly improved version of the analytical result which holds for the relic density in this regime, given in Section 3.4 of \cite{Chu:2011be}, see also \cite{Feng:2008mu,Hambye:2019dwd,Hambye:2019tjt}. Along this ``secluded freeze-out'' regime the relic density not only depends on $m_{DM}$ and $T'/T$ as in the relativistic decoupling case, but also on the annihilation cross section. As in the ordinary freeze-out with $T'=T$, the larger the cross section, the longer the DM remains in thermal equilibrium, the more the remaining DM number density is Boltzmann suppressed. Together with the fact that $\langle \sigma v\rangle$ scale as $1/m_S^4$ (as long as $m_S\gtrsim m_{DM}$), this leads to a relic density which decreases quickly as $m_S/m_{DM}$ decreases, unless this is compensated by a larger value of $T'/T$. This explains the behaviour of the $T'/T$ isocontours in the region where $T'_{dec}/m_{DM}<1$ and $m_S\gtrsim m_{DM}$. For $m_S\simeq 2 m_{DM}$ the annihilation process displays a resonance, as can be seen in Fig.~\ref{fig:Mass_to_mass}.

For even lower values of the mediator mass, when $m_S<m_{DM}$, one enters in the light mediator regime, where the result becomes independent of $m_S$, explaining why the $T'/T$ isocontours are vertical, as in the relativistic decoupling case, but largely shifted with respect to this case. Note that for $m_S<m_{DM}$, the extra $DM DM \rightarrow SS$ annihilation channel opens up, which also shifts the positions of the $T'/T$ vertical lines. In Fig.~\ref{fig:Mass_to_mass} we did not include the effect of this channel, for the sake of showing how the results behave from a same single $DM DM \rightarrow \psi' \bar{\psi'}$ channel everywhere. In this $m_S<m_{DM}$ region the decoupling is non-relativistic too because the $DM DM \rightarrow \psi' \bar{\psi'}$ channel decouples anyway only when DM becomes non-relativistic (since there is no UV mass scale to suppress it in the relativistic regime).

Finally note that the existence of a lighter $\psi'$ particle in the relativistic decoupling scenario does not  cause any problem. Since $\psi'$ also decouples relativistically, if it is stable it also contributes to $\Omega_{DM}$, but by a factor $m_{\psi'}/m_{DM}$ smaller than $\psi_{DM}$. Thus it quickly becomes an irrelevant sub-dominant DM component as soon as it is sizeably lighter than $\psi_{DM}$. If it decays into the SM its effect will be also small because of the reduced number of particles which decay. Similarly, if it is massless, it constitutes extra radiation in the Universe but contributes to the effective number of extra degrees of freedom in a suppressed way, $N_{\psi'}^{eff}\propto g_{\psi'}\cdot(T'/T)^3$, so that it is allowed by the CMB and BBN constraints on this parameter.

Next, we consider the example of a model without any mediator at all. There, since there is no UV mediator mass cutoff to suppress the cross section at a scale above  $m_{DM}$, the decoupling can occur only from a Boltzmann suppression of the number density. Thus one never ends up with a relic density as large as in the relativistic decoupling case. However, we also consider this case because, still, it can lead to a situation where the annihilation cross section decouples at a temperature not far below $m_{DM}$. In this case there is little Boltzmann suppression and one gets a relic density close to the one obtained for a relativistic decoupling. As a result, PeV DM masses are also allowed. To quantify this effect, we consider the well known model where the hidden sector minimally consists of a Dirac fermion coupling to a dark photon
\begin{equation}
{\cal L} \owns - e'\overline{\chi} \gamma^\mu \chi \gamma '_\mu +h.c.\,,
\label{medcouplV}
\end{equation}
where $e'$ is the $U(1)'$ gauge coupling whose associated gauge boson is the dark photon.
Here the relic density is determined by three input parameters: the DM mass $m_{DM}$, the coupling constant $\alpha'\equiv e'^2/4\pi$ and the ratio of the hidden sector to visible sector energy densities at end of inflation, $\rho'/\rho$ (or equivalently the ratio of $T'/T$ once the hidden sector thermalizes). The annihilation process is $DM DM \rightarrow \gamma' \gamma'$. Integrating the corresponding Boltzmann equation for the DM number density, one can determine, as a function of $m_{DM}$ and $\alpha'$, what is the value of  $T'/T$ which leads to the observed relic density, $\Omega_{obs}=0.26$ (imposing that $\Gamma/H>1$ for at least one value of $T'$ so that the annihilation process approximately thermalizes at, at least, one temperature). Assuming this value of $T'/T$, one can also compute as a function of $m_{DM}$ and $\alpha'$ what is the value, ``$\Omega_{rel}$'',  of $\Omega_{DM}$ one would have obtained if DM had decoupled when still relativistic, simply using Eqs.~(\ref{nDMrelativistic}) and (\ref{OmegaDMrelativistic}). 

One can check numerically that ratios $r\equiv \Omega_{obs}/\Omega_{rel}$ as large as 0.9 can be obtained  in the mass range between GeV and few PeV. This means that, even if there is no heavy mediator, one can still get relic densities, and hence DM masses, approximately as high as in the relativistic decoupling scenario. Of course these maximum values of $r$ and $m_{DM}$ are obtained for a single value of $\alpha'$ which is just large enough for the annihilation to approximately thermalize and just small enough not to cause too much Boltzmann suppression. However, they show the range of possibilities for the case without heavy mediator. As soon as one considers smaller values of $m_{DM}$, for example few hundreds of TeV, there is a whole range of possible values of $\alpha'$. In this case one lies already well within the "secluded freeze-out" regime which has been recently studied for this model in \cite{Hambye:2019tjt}, see Fig.~6 of this reference, and, here too, it is a good approximation to use the analytic results obtained for the relic density in Section 3.4 of \cite{Chu:2011be}.\footnote{Note that in \cite{Berlin:2016vnh} it is briefly mentioned, on the basis of the same usual DM model with light dark photons, that for $T'/T\ll 1$, we could get in principle the observed relic density for ``very large DM masses'' with in addition no overclosure of the Universe by the particles into which DM annihilates in the hidden sector.} These have been obtained from plugging in the annihilation rate the non-relativistic form of the DM number density.

All of the above assumes no connector between the visible and hidden sector. For both models above there are nevertheless possibilities of connections through a neutrino portal, ${\cal L}\owns -y_{portal} \overline{\psi}_{DM} L H$, and kinetic mixing portal, ${\cal L}\owns -\frac{1}{2}\, \epsilon \,F_{\mu\nu}^Y F'^{\mu\nu}$, respectively. A Higgs portal communication, ${\cal L}\owns -\lambda_m S^2 H^\dagger H$ is also possible in the first model, and even in the second model if for instance the dark photon gets a mass from the Brout-Englert-Higgs mechanism. The assumption here is that these portals are so tiny that the two sectors do not thermalize. For example, imposing that the portal does not change the relic density by more than, say, $10\%$,  requires $\epsilon\lesssim 10^{-14}$ for $m_{DM}\simeq 1$ PeV and $\alpha '\simeq 0.1$.

Note, however, that such small portals do not necessarily mean that these models could not be tested in ways other than gravitational. For instance, in the first model the portal induces a decay which, in order to be slow enough, requires anyway values of the portal very far below the values which could thermalize both sectors. If the portal is non vanishing, the decay proceeds into $l^\pm W^\mp$, $\nu h$ and $\nu Z$ and leads to a flux of neutrinos which could potentially be the origin of the 6-years IceCube HESE neutrino events at the PeV scale, while also inducing a flux of $\gamma$-rays which is small enough not to be excluded by Fermi-LAT data \cite{Sui:2018bbh,Bhattacharya:2019ucd}. From the fits of IceCube data which have been performed for the $W^+W^-$, $ZZ$, $hh$, $l^+l^-$ and $\nu\bar{\nu}$ decay channels in \cite{Bhattacharya:2019ucd}, it appears that the combination of the channels above which applies for this model leads to a reasonable fit for $m_{DM}\sim 5$~PeV with $\tau_{DM}\sim 5\times 10^{27}$~sec (which is obtained for $y_{portal}\sim3\times10^{-29}$). A lower mass, $m_{DM}\sim 0.6$~PeV, with a slightly smaller lifetime, appears to be possible too \cite{Bhattacharya:2019ucd}.

On the other hand, for the $\gamma'$ model the kinetic mixing portal, as well as a possible Higgs portal, do not lead to any DM decays. In this case, in order to explain the IceCube data, one could think about an annihilation process. However, this would require an annihilation cross section today about 4 orders of magnitude larger than the usual thermal value \cite{Bhattacharya:2019ucd}, while the hot relic scenario requires instead a cross section smaller than the thermal value, as discussed above. Even a large Sommerfeld boost today would not easily lead to a large enough cross section. Nevertheless, this model could leave characteristic signatures in direct detection experiments because in this case the cross section on nucleon is largely enhanced from the fact that it proceeds through the exchange of the light or massless $\gamma'$ \cite{Hambye:2018dpi}.

Finally, note also that it is easy to build models where there is no possibility of renormalizable and gauge invariant portals, especially at the DM scale (even if there could still be possible connections in the UV, especially at the inflation scale). In this case there would be no possibility of tests other than gravitational. However, nothing guarantees that probes other than gravitational necessarily exist. 

In summary, with or without testable portal, if the DM does not thermalize with the SM thermal bath, the relativistic decoupling setup considered above is quite generic and appears to be the most straightforward one could consider. It leads to a relic density determined by only 2 input parameters ($m_{DM}$ and $T'/T$), and works easily for a DM scale all the way from $\sim $~1 keV to few tens of PeV. It constitutes in particular an especially simple scenario to account for the observed relic density for a DM mass scale of order PeV.

\textit{Note added in proof}:
After completion of this work we became
aware of the preprint \cite{Sigurdson:2009uz}, which presents a similar general framework, considering a DM mass up to few TeV.

\section*{Acknowledgments}
We thank Michel Tytgat and Laura Lopez-Honorez for discussions. This work is supported by the F.R.S./FNRS under the Excellence of Science (EoS) project No. 30820817 - be.h ``The H boson gateway to physics beyond the Standard Model'', by the FRIA, by the ``Probing dark matter with neutrinos" ULB-ARC convention and by the IISN convention 4.4503.15.

\pagebreak
\bibliographystyle{apsrev}
\bibliography{bibliography}

\end{document}